\documentclass{article}

\usepackage{arxiv}

\usepackage[utf8]{inputenc} 
\usepackage[T1]{fontenc}    
\usepackage{hyperref}       
\usepackage{url}            
\usepackage{booktabs}       
\usepackage{amsfonts}       
\usepackage{nicefrac}       
\usepackage{microtype}      
\usepackage{lipsum}		
\usepackage{graphicx}
\usepackage[square,numbers]{natbib}
\usepackage{doi}
\usepackage{makecell} 
\usepackage{amsmath} 
\usepackage{longtable} 

\title{Multimodal User Authentication in Smart Environments: Survey of User Attitudes}

\author{{\hspace{1mm}Aishat Aloba}\\
	Department of CISE\\
	University of Florida\\
	Gainesville, Florida, USA\\
	\texttt{aoaloba@ufl.edu} \\
        \And
	{\hspace{1mm}Sarah Morrison-Smith} \\
	Department of Computer Science\\
	Hamilton College\\
	Clinton, New York, USA\\
	\texttt{smorriso@hamilton.edu} \\
        \And
	{\hspace{1mm}Aaliyah Richlen} \\
	Department of CISE\\
	University of Florida\\
	Gainesville, Florida, USA\\
	\texttt{aaliyahrichlen@ufl.edu} \\
        \And
	{\hspace{1mm}Kimberly Suarez} \\
	Department of CISE\\
	University of Florida\\
	Gainesville, Florida, USA\\
	\texttt{kimberly.suarez@ufl.edu} \\
	\And
	{\hspace{1mm}Yu-Peng Chen} \\
	Department of CISE\\
	University of Florida\\
	Gainesville, Florida, USA\\
	\texttt{yupengchen@ufl.edu} \\
        \And
	{\hspace{1mm}Shaghayegh Esmaeili} \\
	Department of CISE\\
	University of Florida\\
	Gainesville, Florida, USA\\
	\texttt{esmaeili@ufl.edu} \\
        \And
	{\hspace{1mm}Damon L. Woodard} \\
	Department of ECE\\
	University of Florida\\
	Gainesville, Florida, USA\\
	\texttt{dwoodard@ece.ufl.edu} \\
	\And
	{\hspace{1mm}Jaime Ruiz} \\
	Department of CISE\\
	University of Florida\\
	Gainesville, Florida, USA\\
	\texttt{jaime.ruiz@ufl.edu} \\
        \And
        {\hspace{1mm}Lisa Anthony} \\
	Department of CISE\\
	University of Florida\\
	Gainesville, Florida, USA\\
	\texttt{lanthony@cise.ufl.edu} \\
}

\date{}

\hypersetup{
pdftitle={Multimodal User Authentication in Smart Environments: Survey of User Attitudes},
pdfsubject={Biometrics},
pdfauthor={Aishat Aloba, Yu-Peng Chen},
pdfkeywords={Privacy, Security, Smart Environments},
}

\begin{document}
\maketitle

\begin{abstract}
As users shift from interacting actively with devices with screens to interacting seamlessly with smart environments, novel models of user authentication will be needed to maintain the security and privacy of user data. To understand users’ attitudes toward new models of authentication (e.g., voice recognition), we surveyed 117 Amazon Turk workers and 43 computer science students about their authentication preferences, in contexts when others are present and different usability metrics. Our users placed less trust in natural authentication modalities (e.g., body gestures) than traditional modalities (e.g., passwords) due to concerns about accuracy or security. Users were also not as willing to use natural authentication modalities except in the presence of people they trust due to risk of exposure and feelings of awkwardness. We discuss the implications for designing natural multimodal authentication and explore the design space around users’ current mental models for the future of secure and usable smart technology.
\end{abstract}


\section{Introduction}
The array of smart technologies that users interact with in their daily lives continues to grow, bringing Mark Weiser’s famous vision of ubiquitous computing \cite{Weiser1999} closer to reality every day. Weiser described a world in which computer technologies would be interconnected with each other and integrated seamlessly with the environment and world at large. The smartphone revolution ushered in an era of unprecedented continuous interaction with smart technology, and we are poised on the next frontier as smart environments become feasible. Cook and Das define a “smart environment” as “a small world where all kinds of smart devices are continuously working to make inhabitants’ lives comfortable” \cite{cook2007}. Unlike smartphones, which provide a screen display and generally require active interaction from the user, future smart environments will enable users to interact seamlessly with their immediate surroundings using more natural communication modalities such as voice and gesture \cite{Argyroudis2004}, and will not always provide or require a screen.

There are many open research questions in terms of how comfortable users will be in using natural multimodal interactions in smart environments, especially for certain tasks and contexts when others may be present. How these interactions expose or grant access to user data (i.e., authenticate users) will have a significant impact on potential user acceptance of these systems. Hence, understanding users’ authentication preferences is a critical component of the design process. Previous work in the security and privacy research community has explored questions around users’ authentication preferences and has shown that their preferences depend on factors such as tasks, modalities, and context (i.e., who else is present) \cite{Baldauf2019, Buschek2016}. However, this work has primarily been done in the context of use of smartphones or other devices linked directly to a user personally, and as such may not generalize to the more open nature of smart environments. We investigated the following central research questions:
\begin{itemize}
	\item[\textbf{RQ1}] What are user preferences regarding using natural input modalities for authentication in smart environments?
        \begin{itemize}
            \item[\textbf{RQ1.1}] Are users’ authentication preferences similar or different for traditional authentication modalities (e.g., passwords) versus natural authentication modalities (e.g., body movements)?
            \item[\textbf{RQ1.2}] Do users’ authentication preferences for smart environments change depending on the modalities or combination of modalities, the context of who else is present in the environment, or the population age group of users?
        \end{itemize}
\end{itemize}
To answer these questions, we deployed a comprehensive survey asking users about authentication preferences. The design of our survey questions did not assume any specific system functionality or the availability of interactive screens. First, we asked respondents to report their preferences towards authentication in general for different tasks (Table~\ref{tab:table1}). The goal of this question was to establish what categories of tasks users want to authenticate for, so that we could then use these categories of tasks as prompts in later questions to understand users’ attitudes toward multimodal authentication. Next, for each category of tasks for which respondents mentioned they would prefer to authenticate, we asked them to rate their preferences for specific authentication modalities. We also asked respondents to rate their preferences for these modalities in different contexts (e.g., alone, with a significant other, with coworkers, etc.), and along various usability dimensions: ease of use, comfort, naturalness/intuitiveness, and security \cite{Khan2019}. For the modalities, we selected some common traditional authentication modalities that users would already have experienced (e.g., fingerprint scanner), and some novel natural modalities enabled by multimodal interaction in smart environments (e.g., voice recognition). Lastly, we asked respondents to rate their preferences for different combinations of modalities (e.g., passwords and biometrics). We deployed the survey on Amazon Mechanical Turk (117 responses) and with graduate and undergraduate students in our computer science program (43 responses). We investigated both populations because prior work has found that the security and privacy experiences of mTurkers are representative of the general U.S. population \cite{Redmiles2019}, but also that younger generations (i.e., < 39-year-olds) are less likely to be concerned about security and privacy compared to older generations \cite{Pereira2017}. By using this mixed sampling approach, we can understand key differences in authentication preferences between these populations. We analyzed responses to closed survey questions quantitatively using statistical tests and open survey questions qualitatively using inductive coding \cite{McCracken1988}.

We found that users place less trust in natural authentication modalities compared to traditional authentication modalities. Our survey respondents generally viewed natural authentication modalities to be not as accurate or secure as traditional authentication modalities, for example fingerprint scanners, so their ratings of these modalities were significantly lower compared to the traditional modalities. Users reported that they would rather combine natural input modalities for two-factor authentication than solely using one of the modalities for authentication. We also found that compared to traditional authentication modalities, users are not as willing to use natural authentication modalities except in the presence of people they trust. Respondents’ ratings of authentication modalities in the presence of others were dependent on the intersection between their trust in those present, comfort, and the risk of exposure of authenticating in a particular modality. They viewed natural modalities as awkward to use and susceptible to observability attacks in the presence of people they distrust. From our findings, we discuss the implications for designing natural multimodal authentication for smart environments and present an analysis and exploration of the design space based on users’ current mental models for the future of secure and usable smart technology.

The contributions of this paper include (a) new knowledge about user preferences with regards to natural authentication modalities in smart environments, particularly in comparison to traditional authentication modalities, especially in different privacy contexts; (b) an understanding of key differences between the general population (MTurkers) and younger generations (CS students) with respect to authentication attitudes; and (c) a characterization of the design space of multimodal authentication for secure and usable smart environments, including novel design recommendations. Our work will inform the future design of multimodal authentication for interaction with smart technology.
\begin{table}
	\caption{List of tasks covered in the survey questions.}
        \label{tab:table1}
	\centering
	\begin{tabular}{lll}
		\toprule
		\cmidrule(r){1-2}
		Task Category & Example tasks \\
		\midrule
		Calendar & \makecell[l]{Check an event time \\ Delete an event time} \\[0.4cm]
		Email/Messages & \makecell[l]{Create email/messages \\ Checking emails} \\[0.4cm]
		Online Banking & \makecell[l]{Transfer money \\ Deposit money} \\[0.4cm]
            Photos & \makecell[l]{Share photos \\ View photos} \\[0.4cm]
            Contacts & \makecell[l]{Share contacts \\ Delete contacts} \\[0.4cm]
            Purchases & \makecell[l]{Add Item to cart \\ Add payment information} \\[0.4cm]
            Access online info & \makecell[l]{Look up online articles \\ Download information} \\
		\bottomrule
	\end{tabular}
\end{table}
\section{RELATED WORK}
We review relevant prior work on user attitudes toward privacy and security in two main areas: (1) authentication techniques; and (2) data privacy and monitoring.
\subsection{Authentication Techniques and User Attitudes}
Authentication is the process of demonstrating that a user is the right person to access a system or a device (also called an authentication target) using some kind of authenticator. An authenticator is “a way to provide evidence that we are the right people to unlock the restricted resources in our lives” \cite{Mare2016}. Authenticators can be knowledge-based (e.g., passwords), token-based (e.g., smart key fob), or biometrics-based (e.g., fingerprints) \cite{Egelman2014}. For example, smartphones and personal computers, as authentication targets, store personal information about the user that can pose security risks if accessed by an unauthorized person. To protect the information stored on these devices, researchers and designers have relied on a lock/unlock model: an ‘all or nothing’ approach in which every task is behind a lock (i.e., an authentication such as a PIN). Users must authenticate for every task, and once unlocked, no authentication is required for further tasks within the same session \cite{Buschek2016, Egelman2014}.

However, prior work has found that users are often frustrated with the notion of having to authenticate for every interaction, especially when accessing applications they believe contain no personal information \cite{Egelman2014}. For example, Hayashi et al. \cite{Hayashi2012} found that participants in their study wanted to authenticate when accessing emails and purchases, which often include personal information, but not when accessing utility or productivity applications (e.g., calculator, calendar). Buschek et al. \cite{Buschek2016} evaluated the type of content users accessed using the SnapApp lock screen, an application that grants users access to content on their smartphones without authentication within a time limit (30s). The authors found that users employed the system for content they perceived as insensitive (e.g., calendar and messages), but rarely used the system for content they perceived as highly sensitive, such as email and shopping.

Contextual factors such as data sensitivity, location, frequency of sharing, and users’ perception of risk also have been shown to affect the type of authentication methods users are comfortable with using. Baldauf et al. \cite{Baldauf2019} compared users’ perception of comfort and suitability of four different authentication modalities (face recognition, fingerprint scanner, NFC ring, and PIN) in four mobile contexts (moving while alone, moving in the presence of others, sitting while alone, and sitting in the presence of others). The authors found that users preferred face recognition in spite of the mobile context. Harbach et al. \cite{Harbach2016} and Buschek et al. \cite{Buschek2016} found that users perceive public spaces, such as the gym or public transports, as more prone to security risks than private spaces, such as their homes or cars. Consequently, users are less willing to authenticate in public spaces, which is likely due to the presence of untrusted or unfamiliar people (e.g., strangers).

Our work adds new insight into the question of user attitudes toward authentication in new contexts and for new authentication targets, including multimodal interactions and smart environments.
\subsection{Data Privacy and User Attitudes}
Prior work has also investigated user attitudes toward the increasing amounts of personal data being stored on users’ devices and perceptions of security and risk associated with this functionality. For example, Chin et al. \cite{Chin2012} found that users are more willing to access personal information on their laptops compared to their smartphones due to the security risks associated with smartphone use. Egelman et al. \cite{Egelman2014} noted that fear of theft, data loss, and mistrust are common risks associated with smartphones. Adding additional complexity, smart environment technologies often interconnect with other devices for data exchange and autonomous computation \cite{Silverio-Fernandez2018}. Thus, interaction with smart environment technologies can lead to higher amounts of sensitive data being collected about users in one central place, such as information about users’ lifestyles (e.g., their purchase preferences) \cite{Montanari2016, Prange2019, Yang2017}. The sensitivity of the data collected raises important concerns about the risks associated with data exposure \cite{Montanari2016}, such as information misuse and leakage \cite{Yang2017}, privacy intrusion \cite{Yang2017}, hacking \cite{Yang2017}, and physical risks in the environment (e.g., burglary \cite{Rentto2003, Yang2017}).

Some prior work has investigated users’ perception of risk associated with using emerging smart environment technologies that can track so much of their own data \cite{Singh2018, Yang2017}. In fact, users’ perceptions of security and privacy risks have a profound impact on their willingness to adopt smart home technologies \cite{Singh2018}; higher perceived risk results in negative attitudes towards smart environment technologies. Users especially tend to perceive risks associated with monitoring of their private activities \cite{Hansen2008}, especially related to the use of cameras \cite{Singh2018}. Furthermore, users have privacy concerns about user identification within a smart environment, particularly with regards to who will have access to the data \cite{Marikyan2019}. Notably, these attitudes are not universal across age groups. Older adults (i.e., adults aged 36 to 70 years old) are more accepting of monitoring and sharing data, especially with their doctors and caregivers \cite{Hansen2008}, and prioritize personal physical security \cite{Park2017}. In contrast, young adults (i.e., adults aged 18 to 35) are more reluctant to share information \cite{Hansen2008} and identify personal data privacy as being of great importance \cite{Park2017}. Prior work further noted that users raise privacy concerns about smart environment technologies when the functionality of the technology (e.g., the type of data that it collects) does not match their pre-existing mental models about the purpose of the technology \cite{Montanari2016}.

Our work aims to examine user attitudes, especially across user populations, specifically toward the act of granting and denying access to users in new contexts like smart environments through natural multimodal authentication.
\section{METHOD}
We designed a 196-question survey covering a range of topics, including: demographics of the respondents, their experience with smart technologies and authentication, authentication preferences for a range of specific scenarios (defined by tasks and context of others present), and ratings of perceived usability and security of various authentication modalities. Details of the survey structure, organization, and questions are provided here. The full survey is attached to this paper as an Appendix.
\subsection{Survey Design}
\subsubsection{Demographics}
We asked survey respondents to identify their gender, age, highest level of education completed, occupation, income level, experience with specific technologies and languages spoken. We also asked respondents to indicate which authentication modalities they had used in the past: type a password, voice recognition, make a body movement, fingerprint scanner, face recognition, and other (with a space to fill in details). For each of their selections, we asked them to identify how long ago it was since they last authenticated using that modality, and to rate how often they have used that modality (e.g., “multiple times a day” to “rarely”). Next, we asked them if they had ever had an account compromised, and if so, what method(s) of authentication protected that account and how they thought the attacker gained access. Finally, we asked them if they had ever forgotten a password or other authentication token, and if so, how had they recovered access.
\subsubsection{Perceptions of Authentication Modalities and Context}
The next section of the survey focused on identifying users’ preferences regarding multimodal authentication. First, we identified tasks for which users would prefer a system to require authentication prior to performing an action, in order to establish what categories of tasks to ask users about in later questions. We selected seven categories of tasks, each of which had three to four related tasks, based on prior work \cite{Buschek2016, Melicher2016} and the authors’ expert mental models (Table~\ref{tab:table1}). Using the task categories, we asked respondents to rate how likely they would be to accept using specific authentication modalities to verify their identity and how likely they would be to use different modalities to authenticate their identity when alone or in the presence of specific individuals or groups of people. Based on prior work \cite{Harbach2016}, we selected the following as the individual/group(s) respondents should consider that they might be with: alone, significant other, roommate, close friends, acquaintances, family, co-workers, and general public. We also asked respondents to rate how (1) easy, (2) comfortable, (3) natural/intuitive, and (4) secure they felt each authentication modality was, and to rate how likely they thought they would be to use each possible combination of authentication modalities in a two-factor set-up. For example, we asked how likely they would be to use a typed password with a biometric authentication method (e.g., fingerprint scanner, face recognition), and repeated these questions for all possible combinations of methods. We included methods in this list currently used for authentication and/or two-factor authentication: biometric, make a body movement, phone call, text message, voice recognition, and type a password. All ratings were done using a visual analog scale (VAS) from 1 (very unlikely) to 5 (very likely). We also included open-ended questions to ask respondents to briefly explain the reasons for their highest and lowest ratings for each authentication modality or combination of modalities and context.
\subsubsection{Attention Check}
Finally, based on prior literature in qualitative user surveys \cite{Cherapau2019}, we included an “attention check” question at the end of the survey which instructed the respondent to mark a specific answer to show they had fully read all questions. This attention check question helped identify respondents that were fully engaged in participating in the survey, answering the questions with thoughtful responses. Survey responses from respondents who failed the attention check were excluded from analysis.
\subsection{Participants and Recruitment}
We deployed the survey on Amazon Mechanical Turk (MTurk), and also offered it to graduate and undergraduate students in our department’s computer science (CS) classes. MTurkers were paid \$8 for submitting the survey, and CS students earned 1 point of extra credit in their class (not to exceed 2\% of the overall grade) for submitting the survey. The survey took approximately 70 minutes to complete on average (mean: 68.9 minutes, median: 40 minutes, min: 8.67 minutes, max: 40.7 hours, stdev: 201.4 minutes). A total of 147 MTurkers and 83 CS students responded to the survey. We excluded 30 MTurk respondents and 40 CS student respondents who either submitted an incomplete survey or failed the “attention check” question at the end of the survey. Thus, our analysis is based on 117 MTurkers and 43 CS students.
\subsection{Survey Analysis}
Respondents’ answers to survey questions provided valuable insight into their current preferences and experiences using authentication modalities. Survey questions included both closed-ended survey responses, which we analyzed quantitatively using statistical tests, and open-ended survey responses, which we analyzed qualitatively using open coding.
\subsubsection{Quantitative Analysis}
We analyzed all closed-ended survey responses quantitatively using repeated measures analysis of variance (ANOVA). Prior to the ANOVA analysis, we first excluded all missing data resulting from the design of our survey (i.e., questions some participants were not asked). In order to conform to ethical standards specified by our Institutional Review Board (IRB), respondents could skip questions that they did not wish to answer, so we also excluded blank responses from respondents. None of our data satisfied the requirements for normality, so we followed recommendations established in prior work \cite{Conover2012} and transformed the data using a rank transformation prior to the ANOVAs. All post hoc analysis was done using Tukey’s method.
\subsubsection{Qualitative Analysis}
We analyzed responses to each of the open-ended questions on our survey separately using qualitative coding \cite{McCracken1988}, in order to determine the categories, relationships, and assumptions that informed our participants’ views regarding authentication \cite{McCracken1988}. We followed a two-stage process to code the survey responses.

In the first phase, similar to methods followed in prior work \cite{Gorden1998}, four researchers independently coded responses from the same 10\% of the MTurk participants and CS student participants for each coding dimension (a coding dimension corresponded to each of the open-ended questions) to create codes. A code is a word or short phrase that captures the underlying meaning behind a portion of language-based data \cite{JohnnyS2009}. Then, the researchers came together to combine and refine the codes for each coding dimension. A survey response could be assigned multiple codes. The purpose of this phase was to generate a codebook. Our full codebook is attached to this paper as an Appendix. In the second phase, three of the researchers were assigned dimensions to code in a counterbalanced manner, such that each dimension was coded independently by two researchers. In total, our data included 7360 total open-ended responses across all 10 open-ended questions (dimensions). Once all the dimensions had been coded, we evaluated the agreement between coder pairs for each dimension using Inter-Rater Reliability (IRR), which is the degree of agreement between raters when accounting for agreement due to random chance \cite{Gwet2010}. We measured IRR using Cohen’s Kappa ($\kappa$), which ranges from -1 (perfect disagreement) to 1 (perfect agreement) \cite{Hallgren2012}. The average Cohen’s kappa across all dimensions and coders was 0.766 (min: 0.661, max: 0.95, SD: 0.083). This value corresponds to \textbf{substantial} agreement \cite{Maghoumi2019}. Next, all three researchers met to resolve any coding disagreements that may have arisen due to subjective judgments about the data and conceptual misunderstandings about codes in the codebook. Once all coding disagreements were resolved, researchers came to a consensus on all responses, allowing for further analysis of themes, reported below.
\section{SURVEY RESULTS}
We organize the presentation of our survey results around the related variables (i.e., modality, setting, and population) that we asked about in the survey questions. Please note that although we did include population in all our ANOVA models, we only present findings related to population in section 5.4 to help streamline the presentation of our results. In all sections, we use the qualitatively coded responses to explain trends and effects seen in the quantitative data.
\subsection{Participants}
The demographic profiles of the two samples covered by our survey are as follows:
\paragraph{MTurkers (N = 117):}
average age 38.67 years (range: 22 to 65 yrs, SD: 9.89 yrs); 72 male, 44 female, 1 transgender. Occupations covered nearly all sectors, including financial, customer service and retail, artists/musicians, marketing, office work, IT and computer technicians, automotive, medical, food services, teaching, security, and homemaker or self-employment. About half of the respondents (53) had an account compromised in the past. Nearly all (106) had forgotten a password or other authentication token before.
\paragraph{CS Students (N = 43):}
average age 21.05 years (range: 18 to 32 yrs, SD: 2.71 yrs); 31 male, 12 female. About 60\% (26) had never had an account compromised before. All respondents (43) indicated they had forgotten a password or other authentication token at some point in their past.
\subsection{Effect of Modality on User Attitudes toward Authentication}
To understand user preferences toward various modalities for authentication, we analyzed users’ responses to survey questions to identify which modalities users would prefer for authentication in general, how usable they think each of these modalities is for authentication, and which modalities users would prefer for two-factor authentication. We first analyzed responses to survey questions about the categories of tasks users would prefer to authenticate for, since we used them in later questions when asking about users’ modality preferences.

We calculated the number of tasks for which respondents selected that they would prefer to require authentication and divided it by the total number of tasks within that task category, which we call the task authentication ratio. We performed a one-way repeated measures ANOVA on the task authentication ratio with a within-subjects factor of category (accessing online information, calendar, contacts, email, making a purchase, online banking, photos). To account for the repeated nature of categories, we included respondent ID as a random factor. We found a significant main effect of category ($F_{6,948} = 62.05, p < 0.0001$). Post hoc analysis revealed that there was a significant difference in all category pairs except between Making a Purchase/Photos (M = 0.73 [SD = 0.30] vs. 0.70 [0.35]), Making a Purchase/Contacts (M = 0.73 [0.30] vs. 0.66 [0.37]), and Online Banking/Emails+Messages (M = 0.88 [0.24] vs. 0.85 [0.29]). Users regarded categories involving financial information as needing the most security and accessing online information as needing the least security. We found that while almost all respondents preferred to require authentication for Online banking (98\%) and Making a Purchase (96\%) respectively, a smaller number of respondents wanted to require authentication for Accessing online information (78\%). These results align with the expected sensitivity of data associated with financial tasks versus other tasks. Our qualitative results also support this idea that users consider financial information to be sensitive (“\textit{banking information is very sensitive, and I would not want my privacy invaded}” [m103]\footnote{Anonymous respondent IDs are given to provide context for quotes: m\# indicates the respondent was from Mechanical Turk; s\# indicates a CS student.}). Users’ mental models regarding authentication centered around wanting to exert \textbf{control} over who had access to their information (“\textit{[I] would not want someone else to have any access whatsoever to my emails.}” [m28]), the \textbf{privacy} of the information being accessed (“\textit{my events are private...}” [m109]), and the \textbf{risk} associated with unauthorized access (“\textit{I don't want someone to delete events that might be important to me.}” [m10]). Overall, our findings show that even for the least sensitive task (accessing online information), about 80\% of our respondents still wanted to authenticate for that task. Based on these findings, one may suggest that designers should require authentication for all tasks. However, our findings also showed that respondents expressed their annoyance for having to authenticate for tasks that they think should not require authentication (“\textit{There should be no authentication for any of these. I want unrestricted access to my photos with an uninterrupted user experience.}” [m86]). Therefore, we suggest that designers should consider adaptive, personalized, or customizable authentication models to allow users to decide.
\subsubsection{General Modality Preference}\label{sec:section5.2.1}
For each task category, we asked security-minded participants (respondents who had preferred to require authentication for that category) about their preferences for using different authentication modalities (type a password, voice recognition, make a body movement, fingerprint scanner, face recognition) on a scale of 1 to 5. We calculated the \textit{average rating} (per person) for each authentication modality across all task categories; we excluded 2.13\% of the data due to missing responses (e.g., respondents who did not prefer to require any authentication for a given task category). Users rated type a password the highest (M = 4.46 [SD = 1.04]), followed by fingerprint scanners (M = 3.75 [1.57]), face recognition (M = 2.94 [1.62]), voice recognition (M = 2.58 [1.58]), and make a body movement in front of a camera (M = 1.76 [1.18]). A 2-way repeated measures ANOVA on \textit{average rating} with a within-subjects factor of \textit{authentication modality} and a between-subjects factor of \textit{population} (MTurkers, CS students), with \textit{respondent ID} as a random factor found a significant main effect of \textit{authentication modality} ($F_{4,619.7} = 116.01, p < 0.0001$). Post-hoc analysis revealed a significant difference between all authentication modality pairs except between type a password and fingerprint scanner. Security-minded respondents rated typed passwords and fingerprint scanner significantly higher than other modalities, especially “natural” modalities, such as voice and gesture, which were rated the lowest. We further investigated whether users’ prior experience using a modality for authentication influenced their ratings by analyzing the relationship between respondents’ answers to the survey question about the authentication modalities they previously used and the \textit{average rating} of each authentication modality. The authentication modality that has been used by most respondents was type a password (100\%), followed by fingerprint scanner (72.5\%), face recognition (40\%), voice recognition (27.5\%), and make a body movement (0\%). We used a Chi-squared test of independence to test the hypothesis that the distribution of respondents’ previous modality experience and average modality ratings were independent; it showed a significant relationship ($\chi^2$(4) = 441.96, p < 0.01). This finding suggests that \textbf{previous experience with a modality is a driving factor behind respondents’ ratings of each modality’s suitability for authentication}.
\subsubsection{Usability}\label{sec:section5.2.2}
To further understand users’ preference for different authentication modalities, we investigated users’ perception of the usability of these modalities with respect to four dimensions: (1) \textit{ease of use}, (2) \textit{comfort}, (3) \textit{naturalness/intuitiveness}, and (4) \textit{security}. For each dimension, we asked all survey respondents to tell us more about their perceptions of each authentication modality. We performed a 2-way repeated measures ANOVA on rating with a between-subjects factor of \textit{population} (MTurkers, CS student) and a within-subjects factor of \textit{authentication modality} and \textit{respondent ID} as a random factor (we excluded 2.38\%, 2.00\%, 2.25\%, and 2.38\% of the data for ease, comfort, naturalness, and security respectively, due to missing responses). We found a significant main effect of \textit{authentication modality} across all dimensions: ease of use ($F_{4,619.8} = 68.06, p < 0.0001$), comfort ($F_{4,619.4} = 176.79, p < 0.0001$), naturalness/intuitiveness ($F_{4,618.3} = 118.95, p < 0.0001$), and security ($F_{4,619.7} = 151.79, p < 0.0001$). Post-hoc analysis revealed that users rated \textbf{fingerprint scanner significantly higher than face recognition, voice recognition, and make a body movement across all dimensions, but only significantly higher than type a password with respect to how easy and secure it is to use} (Table~\ref{tab:table2}). Respondents’ rationale indicated they felt that fingerprint scanner requires no effort to use: “\textit{Does not get much easier than putting your finger on a button.}” [m107] and that the uniqueness of the finger makes the modality secure: “\textit{No one has your fingerprint; they cannot be duplicated. Therefore, this is a very secure way of keeping your identity safe.}” [m64]. Furthermore, the years of experience users have had using the modality also factored into their rationale, especially with respect to the similarity between their ratings of fingerprint scanners and type a password for both the comfort and natural dimensions. For example, participant m15 stated: “\textit{I use a fingerprint scanner a lot to unlock my phone and it has become second nature at this point.}” Similarly for type a password, participant m91 stated: “\textit{Have been typing passwords for many years. Has pretty much become second nature.}”

Post-hoc analysis further revealed \textbf{no significant difference between users’ ratings for face recognition and type a password with respect to how easy and secure they are to use}. Respondents’ rationale indicated that memorability factored heavily into how easy they perceive these modalities to be. For passwords, participant m47 stated: “\textit{Typing a password is easy to do so long as you remember what the password is,}” and the same participant stated for face recognition that: “\textit{Facial recognition is also very easy to use since you do not need to remember or memorize anything other than looking into a camera as you would naturally.}” Although respondents rated both type a password and face recognition as relatively secure, their rationales noted that they felt that these modalities can still be vulnerable to security risks, especially if exploited by malicious users (e.g., hackers). Regarding typing passwords, respondent m107 stated: “\textit{I suppose passwords can always be hacked but they are pretty secure for the most part if proper precautions are followed.}” For face recognition, another respondent stated: “\textit{I think it's pretty secure, but it feels like it's still hackable (e.g., using photos or triggered by camera when you did not intend to).}” [s42].

Across all dimensions, \textbf{natural modalities (i.e., voice and gesture) were rated the lowest} (Table~\ref{tab:table2}); post-hoc analysis revealed that voice recognition and make a body movement in front of a camera were rated significantly lower than all other modalities across all dimensions. Respondents often noted in their rationale that these modalities make (or would make) them uncomfortable. For voice recognition, participant m114 stated “\textit{I don’t feel comfortable talking to my electronics. It seems odd.}” For make a body movement, participant s25 stated: “\textit{I would feel awkward making random [body movements] in front of a camera.}”; m18 also mentioned that “\textit{[it] feels really uncomfortable and forced and gimmicky to me.}” This discomfort is likely because of respondents’ lack of prior experience using these modalities for authentication: “\textit{Have never used body movement, it might be a little awkward.}” [m91]; “\textit{I don't have that much experience with voice recognition as a form of verification, so it still feels a little awkward at this point.}” [m15]. The risk associated with using these modalities for authentication was also a driving factor behind respondents’ rationales. They generally felt that both of these modalities were easy to compromise: respondents stated that making body movements for authentication is “\textit{insecure because the movements can be copied super easy and a lot of times, they are pretty generic so anyone could remember them,}” [m113] while “\textit{voice can be copied or recorded so it makes it less secure.}” [m108]. Interestingly, even though voice and body movements are both natural communication modalities \cite{BERNARDIS2006178}, \textbf{make a body movement in front of a camera was rated significantly lower than voice recognition} across all dimensions. Respondents were generally unwilling to use body movements for authentication: “\textit{I would never use this as a way to verify my identity.}” [m82] and felt that it was difficult to use: “\textit{It requires more efforts than other authentication methods to move around your body.}” [s42].
\begin{table}
	\caption{Average ratings of authentication modality by usability dimension (standard deviations in brackets). Superscripts correspond to the row number of each modality that is significantly different from the modality in the row at p < 0.001 level.}
        \label{tab:table2}
	\centering
	\begin{tabular}{llllll}
		\toprule
		& \textbf{Modality} & \textbf{Ease of Use} & \textbf{Comfort} & \textbf{Naturalness} & \textbf{Security} \\
		\midrule
		1 & Fingerprint Scanner & $4.56 [0.84]^{2,3,4,5}$ & $4.38 [1.20]^{3,4,5}$ & $4.39 [1.02]^{3,4,5}$ & $4.50 [1.01]^{2,3,4,5}$ \\[0.2cm]
		2 & Type a Password & $4.14 [1.07]^{1,4,5}$ & $4.65 [0.80]^{1,4,5}$ & $4.46 [0.93]^{1,4,5}$ & $4.05 [0.96]^{1,4,5}$ \\[0.2cm]
		3 & Face Recognition & $3.75 [1.30]^{1,4,5}$ & $3.28 [1.48]^{1,2,4,5}$ & $3.39 [1.42]^{1,2,4,5}$ & $3.55 [1.44]^{1,4,5}$ \\[0.2cm]
            4 & Voice Recognition & $3.38 [1.31]^{1,2,3,5}$ & $2.66 [1.43]^{1,2,3,5}$ & $2.90 [1.39]^{1,2,3,5}$ & $2.79 [1.30]^{1,2,3,5}$ \\[0.2cm]
            5 & Make a Body Movement &  $2.54 [1.33]^{1,2,3,4}$ & $1.68 [0.93]^{1,2,3,4}$ & $1.89 [1.09]^{1,2,3,4}$ & $1.57 [0.84]^{1,2,3,4}$ \\
		\bottomrule
	\end{tabular}
\end{table}
\subsubsection{Two-factor Authentication}
So far, our results have only shown users’ preferences for each authentication modality independently (i.e., single-factor authentication). However, the trend in existing smart technologies, such as smart phones and desktop computers, is toward multi-factor authentication, which has been shown to make authentication more secure compared to single-factor authentication \cite{DeFigueiredo2011}. However, users’ preferences toward multi-factor authentication, especially in smart environments, remains unclear. We investigated users’ preferences for different authentication modality combinations for two-factor authentication. We asked respondents to rate how likely they thought they would be to use different pairs of authentication modalities as a two-factor authentication method on a scale of 1 to 5; 2.13\% of the data was excluded due to missing responses for these questions. We performed a 2-way repeated measures ANOVA on \textit{rating} with a between-subjects factor of \textit{population} (MTurkers, CS student) and a within-subjects factor of \textit{two-factor modality} (15 total pairwise combinations of biometrics, make a body movement, phone call, text message, voice recognition, and type a password), and \textit{respondent ID} as a random factor. We found a significant main effect of \textit{two-factor modality} ($F_{14,2164.3} = 47.19, p< 0.0001)$. Post-hoc analysis revealed significant differences between many two-factor modality pairs. We used the post-hoc results to identify two groups of modality pairs (Group 1 and Group 2) such that modality pairs within Group 1 were not rated significantly different from each other but were rated significantly higher than all other modality pairs in Group 2. The groupings are shown in Table~\ref{tab:table3}, ordered from lowest to highest average ratings by the respondents. It is notable that the modality pairs in Group 1 are all combinations that are already being used in current smart devices for two-factor authentication. Therefore, respondents are likely to have experienced these pairs together in the past, which is supported by their rationales: “\textit{Using my regular password and getting a text message or a second authentication is something I do on a fairly regular basis and it doesn't bother me.}” [m74]. This finding suggests that \textbf{users’ prior experience using authentication modalities influences their preferences for these modalities}, which further echoes our findings regarding users’ preferences toward single-factor authentication modalities. All the modality pair combinations involving make a body movement are in Group 2, which continues to show how currently unwilling users are to use natural authentication modalities, such as make a body movement for authentication. Furthermore, respondent ratings for these modality pairs were not significantly different from one another, the only exception being make a body movement+voice recognition. Respondents \textbf{are more willing to combine body movements with voice recognition than combining body movements with other modalities}. From their rationales, we can see that respondents felt that voice recognition might reduce the risk associated with using body movements for authentication: “\textit{A body movement can be easily replicated so pairing it with voice recognition would add an extra layer that is more difficult to reproduce.}” [m47]. It is also noteworthy that modality pair combinations in Group 2 mostly include either phone call or voice recognition, and none of those in Group 1 do. Respondents may have rated these modalities lower because they are not always convenient or appropriate to use. For example, respondents noted they may be unable to answer a call in certain locations or to use voice recognition when unfamiliar people are present in the environment: “\textit{A phone call is not always able to be answered and seems like too much work.}” [m96].

These findings thus imply that designers of future smart environments will need to consider the trade-off between the naturalness of the natural authentication modalities and users’ “trust” in traditional authentication modalities when deciding on the modalities to provide to users for authentication. We elaborate more on how designers can navigate this trade-off in the discussion section.
\begin{table}
	\caption{Groupings of two-factor authentication modality pairs that were rated similarly by survey respondents. Pairs in Group 1 or in Group 2 were not rated significantly different from each other; pairs across groups were.}
        \label{tab:table3}
	\centering
	\begin{tabular}{ll}
		\toprule
		\multicolumn{2}{c}{\textbf{Two-Factor Authentication Pairs}} \\
		\midrule
            \textbf{GROUP 1} & \textbf{GROUP 2} \\
            \midrule
            MEAN: 3.81, SD: 0.03, MIN: 3.78, MAX: 3.84 & MEAN: 2.29, SD: 0.42, MIN: 1.60, MAX: 2.85 \\
		\midrule
		Type a password+text message & Phone call+voice recognition \\
		Biometrics+text message & Biometrics+voice recognition \\
		Type a password+biometrics & Text message+voice recognition \\
             & Type a password+voice recognition \\
             & Biometrics+phone call \\
             & Make a body movement+voice recognition \\
             & Type a password+phone call \\
             & Text message+phone call \\
             & Biometrics+Make a body movement \\
             & Type a password+Make a body movement \\
             & Text message+Make a body movement \\
             & Phone call+Make a body movement \\
		\bottomrule
	\end{tabular}
\end{table}
\subsection{Effect of Context on Users’ Attitudes toward Authentication}
Beyond understanding users’ preferences of authentication modalities generally, we also wanted to know whether the context of who else may be present in the environment where the authentication is occurring would influence users’ willingness to authenticate in that environment and their perception of the suitability of different modalities for authentication.

\begin{figure}
	\centering
        \includegraphics[scale=0.8]{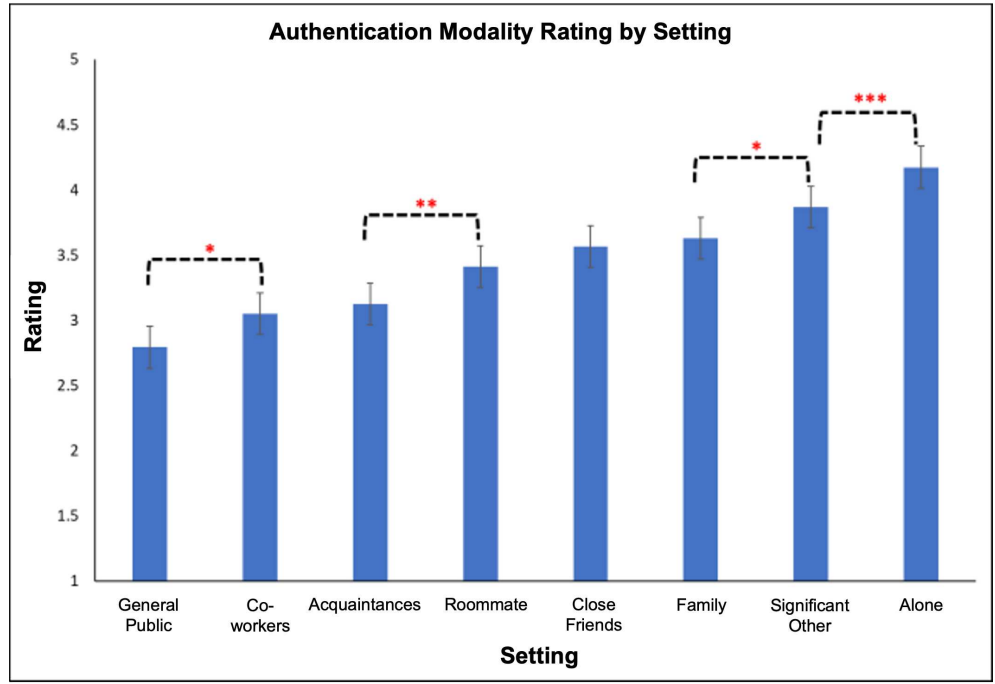}
	\caption{Users’ ratings of authentication modalities by setting, sorted in increasing order of the average rating. Error bars represent 95\% confidence interval. (*) indicates significance at p < 0.05 level, (**) indicates significance at p < 0.01 level, and (***) indicates significance at the p < 0.001 level.}
	\label{fig:fig1}
\end{figure}

We asked all respondents to rate their preferences for using different authentication modalities in the presence of different individuals or groups of people on a scale of 1 to 5; we excluded 1.18\% of the data due to missing responses. We performed a 3-way repeated measures ANOVA on \textit{rating} with within-subjects factors of \textit{authentication modality} and \textit{setting} (alone, significant other, roommate, close friends, acquaintances, family, co-workers, and general public) and a between-subjects factor of \textit{population} (MTurkers, CS students), with \textit{respondent ID} as a random factor. As before, we found a significant main effect of \textit{authentication modality} ($F_{4,632} = 99.37, p < 0.0001$). We also found a significant main effect of \textit{setting} ($F_{7,1118.1} = 91.94, p < 0.0001$). Post-hoc analysis revealed a significant difference between all setting pairs except between co-workers/acquaintances (M = 3.05 [SD = 1.69] vs. 3.12 [1.67]), family/close friends (M = 3.63 [1.56] vs. 3.57 [1.57]), and roommates/close friends (M = 3.41 [1.61] vs. 3.57 [1.57]); see Figure \ref{fig:fig1}. Of particular interest is the similarity in respondents’ preferences to authenticate in the presence of roommates and close friends. This similarity implies that, at least in our sample, users tend to be close friends with their roommates, rather than simply acquaintances.

In general, respondents were also willing to use all authentication methods equally well and at high rates when alone. Respondents noted in their rationales that the risk associated with authenticating when alone is minimal or non-existent: “\textit{alone is always the safest.}” [m108]; “\textit{Always best to do it alone. Don't give anyone a clue or material to help get around the authentication method.}” [m111]. However, \textbf{for settings that included the presence of others, respondents were more willing to authenticate using a variety of modalities in the presence of their significant other compared to authenticating in the presence of any other person or group of people}. Respondents rated the general public significantly lower than any other person or group of people, generally expressing high trust in their significant other (“\textit{I can trust my significant other with anything.}” [m109]) and high distrust in the general public (“\textit{I would not trust the general public with the opportunity to see my password information.}” [s29]). We also found a significant interaction effect between \textit{authentication modality} and \textit{setting} ($F_{28,4358.2} = 21.60, p < 0.0001$). Respondents’ \textbf{ratings of all modalities except fingerprints were affected by the context of who else was present in the environment}; fingerprints were always rated highly, regardless of who else was in the room. Survey respondents expressed that this modality was secure: “\textit{Using my fingerprint is very secure, no matter who is around. I would be fine using it around just about everyone.}” [m116]. This finding is consistent with our findings regarding usability of fingerprint scanners. On the other hand, respondents’ ratings for make a body movement and voice recognition were significantly higher for settings in which they trusted the people present compared to settings in which they did not trust the people present. Respondents expressed that performing these modalities in the presence of people they distrust would make them uncomfortable (“\textit{I don't feel comfortable making strange gestures in public. I think it's embarrassing}” [m80]) and would increase the risk of exposure (“\textit{I would be concerned that someone I don't trust might attempt to make a recording so they could gain access to my account.}” [m110]).

Our findings thus imply that designers of future smart environments will need to consider making natural authentication modalities more discreet if they want users to authenticate with these modalities when others are present in the environment. We will discuss how designers can make natural modalities more discreet in the discussion section.
\subsection{Effect of Population on User Attitudes toward Authentication}
As previously mentioned, our survey respondents belonged to one of two populations: MTurkers, who are fairly representative of the general U.S. population, and CS students, who represent a younger segment of the population. In this section, we investigate whether any differences in authentication preferences exist between the populations. All results we present in this section will rely on ANOVA models we presented in the previous sections, but here we focus only on the main effect of population and interaction effect between population and the factors we are reporting (e.g., authentication modality).

\begin{figure}
	\centering
        \includegraphics[scale=0.8]{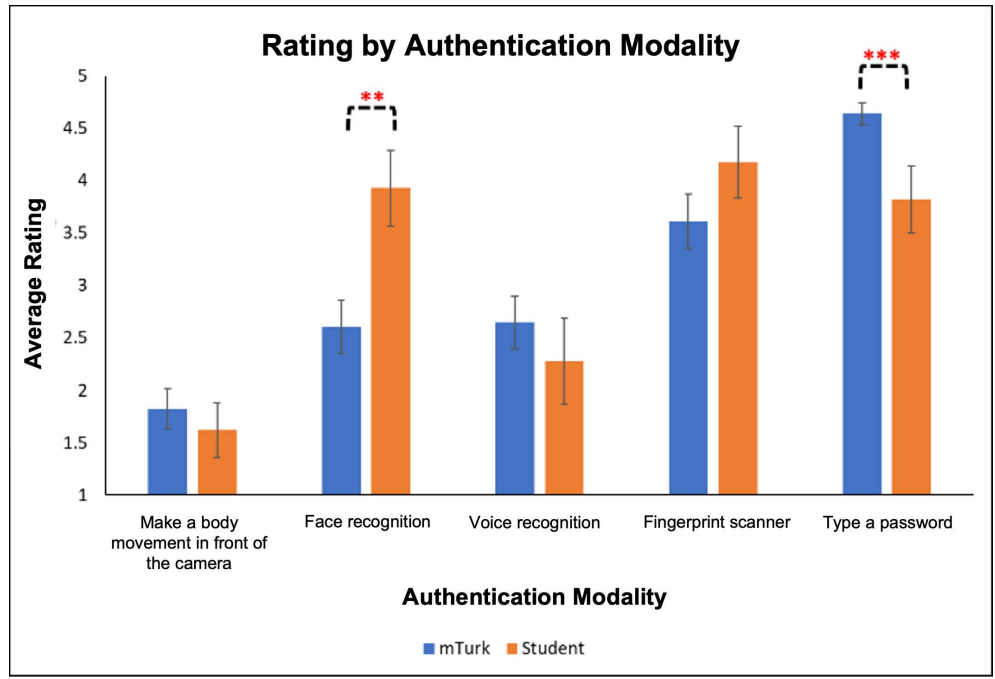}
	\caption{Users’ ratings of authentication modalities by population. Error bars represent 95\% confidence interval. (**) indicates significance at p < 0.01 level, and (***) indicates significance at the p < 0.001 level.}
	\label{fig:fig2}
\end{figure}

To understand whether any key difference exists in MTurkers’ and students’ preference of authentication modalities, we compared both populations’ ratings of different authentication modalities. Based on the 2-way repeated measures ANOVA on average rating with factors authentication modality and population from Section \ref{sec:section5.2.1}, we found no significant main effect of \textit{population} ($F_{1,157.9} = 0.77, n.s.$) (Figure \ref{fig:fig2}) \textbf{MTurkers preferred to use each authentication modality at similar rates as CS students}. However, we found a significant interaction effect between \textit{authentication modality} and \textit{population} ($F_{4,619.7} = 21.48, p < 0.0001$). MTurkers and students differed in their ratings for face recognition, in which students had higher ratings compared to MTurkers. Whereas students focused on the security of face recognition: “\textit{Face recognition is also using unique features of a person’s face so I would not worry about using it in front of people.}” [s42], MTurkers more often expressed distrust in the surveillance aspect of the technology: “\textit{I do not want companies to have a picture of me and track me through cameras.}” [m86]. They also differed in their ratings for type a password, for which MTurkers gave higher ratings compared to students.

We also compared both populations’ usability ratings of authentication modalities along the four usability dimensions (ease, comfort, naturalness, and security) using the ANOVA model from section \ref{sec:section5.2.2} (a 2-way repeated measures ANOVA on average rating with factors population and authentication modality). We found no significant main effect of \textit{population} for any dimension except \textit{ease of use}: \textit{ease} ($F_{1,159.2} = 4.91, p < 0.05$), \textit{comfort} ($F_{1,156.6} = 0.40, n.s.$), \textit{naturalness} ($F_{1,156.3} = 0.57, n.s.$), security ($F_{1,159.3} = 0.02, n.s.$). MTurkers (M = 3.75, SD = 1.33) rated modalities as easier to use on average compared to CS students (M = 3.46, SD = 1.47). Across all usability dimensions, we found a significant interaction effect between \textit{population} and \textit{authentication modality}: \textit{ease} ($F_{4,619.8} = 6.14, p < 0.0001$), \textit{comfort} ($F_{4,619.4} = 15.94, p < 0.0001$), \textit{naturalness} ($F_{4,618.3} = 9.37, p < 0.0001$), and \textit{security} ($F_{4,619.7} = 6.19, p < 0.0001$). MTurkers rated the comfort, naturalness, and security of face recognition lower than students did. mTurkers felt that using face recognition is strange and were generally unwilling to use this modality for authentication: “\textit{It can seem odd to gain access through a face scan.}” [m16]; “\textit{Too bad face recognition technology is so darn bad.}” [m104]. In contrast, MTurkers rated type a password as easier to use than CS students did. While MTurkers often believed that passwords are “\textit{easy enough to remember}” [m17], students sometimes felt that “\textit{It’s a pain to have to remember passwords.}” [s12]. They also rated voice recognition as being more comfortable than CS students did, since students “\textit{…feel awkward talking to a device.}” [s31]. In terms of similarities and differences between user populations for two-factor authentication, we found no significant effect of \textit{population} ($F_{1,156.8} = 0.19, n.s.$), but did find a significant interaction effect between \textit{population} and \textit{two-factor authentication modality} ($F_{14,2164.3} = 2.58, p < 0.01$). MTurkers rated type a password+biometrics lower than CS students did. In their rationales, CS students noted that they felt that using biometrics (e.g., fingerprints) was intuitive and secure: “\textit{Finger scanner makes the most sense and is the quickest way. Also, the most secure.}” [s20]. However, some mTurkers were less willing to use biometrics at all: “\textit{I am not comfortable giving my phone access to my biometric data.}” [m40].

Based on our findings, our younger sample of users and our MTurker sample predominantly differed with respect to their preferences for traditional authentication modalities but were more similar in their (lower) preferences for natural authentication modalities. Although prior work has noted that younger users are more willing to use emerging technologies compared to the general population of users \cite{Czaja2006}, our findings indicate that this finding may not hold true when the emerging technology relies on natural modalities.
\section{DISCUSSION}
Our survey findings revealed the extent to which prior experience, usability, and context/setting influence users’ mental models for how they would prefer to authenticate in smart environments for everyday tasks. We reflect on the implications of our findings for the future design of authentication experiences that meet user expectations. We also use our findings to characterize the design space for building secure and usable smart technology in the future and outline potential future system concepts that meet user expectations.
\subsection{Implications of Main Findings}
In this section, we discuss the implications of our findings on the design of natural multimodal authentication in smart environments.
\subsubsection{Modality Trust vs. Naturalness Trade-Off}
Our findings showed that, despite the potential utility of natural input modalities (voice and gesture) for authentication in smart environments, users place less trust in these modalities because they view them as not as accurate or secure as traditional authentication modalities. In our results, we saw that users’ ratings of an authentication modality were related to their prior experience using that modality, so the more commonly-used traditional authentication modalities were rated higher compared to emerging natural modalities. Based on access and availability alone, the authentication modalities users are currently likely to have the most experience with are type a password and fingerprints, which are the predominant authentication methods on currently widespread devices like smartphones and computers. However, the future of smart technology is likely to bring with it new authentication modalities enabled by different sensors. Smart technology is also likely to move toward “natural” seamless communication modalities, rather than active interaction with keyboards and fingerprint scanners. We are already seeing these types of interactions becoming more common with current smart devices like Google Home and Amazon Echo, which rely on natural human-human communication modalities, such as voice, for interaction and authentication \cite{Hoy2018}. For example, Amazon Alexa performs passive authentication using a \textit{voice profile}, which personalizes users’ interaction experiences using voice recognition \cite{Crist2017}. Therefore, designers of future smart environments that want to use natural input modalities for authentication will need to consider the trade-off between the naturalness of the authentication modality and users’ greater trust in (and experience with) traditional authentication modalities. To bridge this trade-off, designers should provide opportunities for users to gain familiarity using natural input modalities for authentication. As users become more familiar with these modalities, their trust in natural input modalities for authentication will increase. Specifically, designers can explore combining traditional authentication modalities with natural input modalities for two-factor authentication (e.g., during authentication in smart devices). Two-factor authentication approaches have also been shown to be more trusted by users for sensitive tasks (e.g., online accounts) \cite{Reese2019}. By introducing natural input modalities as the second factor, designers are helping users gain more experience using these modalities for authentication, which will increase their trust in these modalities for authentication.
\subsubsection{Social Relationships and Risk Perceptions}
Our findings showed that users’ preferences for authentication modalities in contexts when others are present depends on the intersection between users’ trust in those present, the comfort associated with performing the modality, and the risk of exposure of authenticating in a particular modality. In our results, we saw that survey respondents were mostly unwilling to use natural authentication modalities except in the presence of people they trust. They felt that speaking and gesturing are both actions that are difficult to conceal, so these modalities would be more prone to observability attacks and uncomfortable to perform when others are present. This finding corroborates prior work on gesture acceptability which found that people are more comfortable gesturing in front of people they are familiar with (e.g., partner, friends) versus in front of strangers \cite{Rico2010}. Since smart environments will be deployed in a more open manner in shared settings, rather than being directly linked to one user personally like smartphones, they are likely to be more prone to security and privacy risks \cite{Eiband2017, Prange2019}. Therefore, designers of future smart environments that want to use natural input modalities for authentication should consider ways to make these modalities more discreet to improve likelihood of acceptance. One way to do so is to use user-centric approaches to design authentication interactions that are based directly on users' mental models. For example, designers could use an elicitation study \cite{Good1984, Wobbrock2009} to elicit speech and gesture interactions that users consider comfortable and secure to use in both public and private settings.
\subsection{Design Space: Multimodal User Authentication}
\begin{figure}
	\centering
        \includegraphics[scale=0.5]{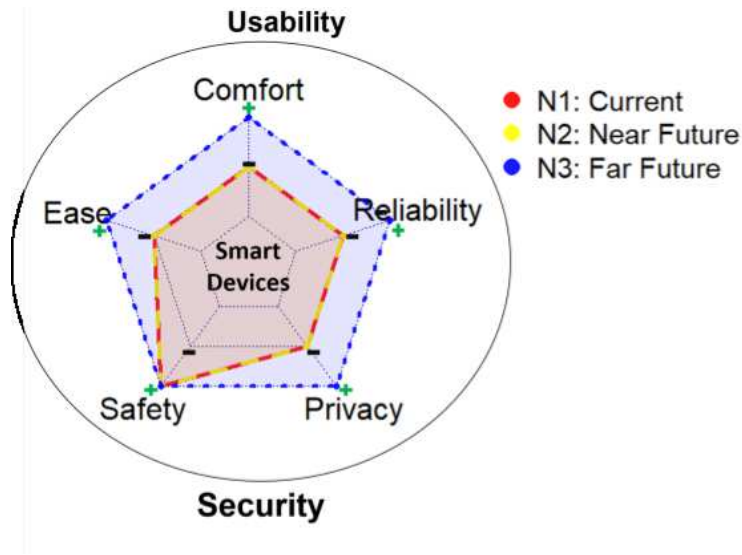}
	\caption{Design Space for authentication in smart devices. Note that although N2 is assigned to \textit{-ease}, this concept is easier to use compared to N1 because of its reliance on only natural authentication modalities.}
	\label{fig:fig3}
\end{figure}

Based upon our findings, especially users’ rationales for their authentication preferences from the open-ended questions on our survey, we characterize the design space for natural multimodal user authentication in smart environments, heavily informed by users’ mental models (Figure \ref{fig:fig3}). From our survey results, we see that users’ mental models regarding authentication center around two main factors: a) usability and b) security. Each of these factors has different facets which are important to a user-centered model of authentication across contexts and modalities. We describe each main factor and their facets here.
\begin{itemize}
    \item \textbf{Usability}: user-perceived usability of authentication modalities.
        \begin{itemize}
            \item \textit{Ease of Use}: amount of effort required to authenticate: users’ mental models expect that authentication should be simple and require no extra effort.
            \item \textit{Comfort}: level of comfort associated with authenticating in a given modality in the presence of others: users’ mental models expect that interaction behaviors required to authenticate in smart devices should not make them feel awkward or embarrassed when others are present.
            \item \textit{Reliability}: degree of accuracy associated with authenticating using a given modality: users’ mental models expect that the modality used for authentication should be accurate in distinguishing authorized users from unauthorized users.
        \end{itemize}
    \item \textbf{Security}: user-perceived risk associated with authentication modalities.
        \begin{itemize}
            \item \textit{Safety}: ease of copying interaction behaviors: users’ mental models expect that authentication interactions should be difficult to spoof and hard to mimic by unauthorized users (e.g., hackers).
            \item \textit{Privacy}: possibility of exposure of interaction behaviors the user performs: users’ mental models expect that they should be able to conceal their authentication interactions from others who may be present to prevent unauthorized access.
        \end{itemize}
\end{itemize}
We next use the lens of our design space analysis to present several new design concepts that meet the user-centered perspective we have presented, organized by expected evolution of future smart technology.
\subsubsection{New Design Concepts}
We propose three new design concepts to inspire further research and design. Each concept is based on our survey findings, and emphasizes \textit{current} technological capabilities, \textit{near future} capabilities, and \textit{far future} capabilities, respectively. Each succeeding concept builds off the preceding one with the end goal being the \textbf{design of natural multimodal authentication that conforms to all facets of our design space}. These are not meant to be comprehensive or complete concepts, but rather to inspire new directions into user-centered usable privacy and security research and design. Figure \ref{fig:fig3} shows a visual representation of the characterization of these concepts using our design space:
\paragraph{Current (N1):}
At the initial stage, our focus is to build users’ trust in natural authentication modalities by taking advantage of the technological capabilities of devices that are currently in use. Since current devices rely on traditional authentication modalities, a system could support a two-factor modality combination using a traditional authentication method, such as fingerprint scanner, as the first factor and a natural authentication modality, such as voice recognition, as the second factor. For example, Elshamy et al. implemented a system that combines voiceprint and fingerprint for authentication by creating a secure dual-biometric system for accessing voicemail \cite{Elshamy2020}. The design concept we propose at this stage will not fully satisfy our users’ goal of being easy to use since, depending on the modalities being combined (e.g., fingerprint scanner and body movements), authentication may require more effort from the user. This system also runs the risk of exposure (i.e., not private) and users may feel awkward using this modality in the presence of others (i.e., not comfortable) because natural modalities are more observable. However, the use of two-factor authentication makes this concept difficult to spoof \cite{DeFigueiredo2011}. Even though the concept does not satisfy most of the facets of our design space, current technologies still largely rely on traditional authentication modalities (e.g., smart phones often use passwords and fingerprint scanners for authentication \cite{Kunda2018}). Therefore, this concept reflects what is currently feasible based on current technology. This concept is also a necessary transition towards increasing users’ trust in natural authentication modalities by increasing their familiarity with these modalities as they use them. Therefore, this concept can be characterized based on our design space as: \textit{– ease – comfort – reliability + safety – privacy}.
\paragraph{Near Future (N2):}
As users continually gain experience combining traditional and natural modalities for authentication (i.e., N1), eventually they will become familiar with using natural modalities. With more familiarity using these natural modalities, users’ trust in them should increase, since our findings showed that familiarity is a major contributing factor to users’ trust in modalities. Based on the current state-of-the-art performance of voice recognition (i.e., \cite{Villalba2020}) and gesture recognition (i.e., \cite{Maghoumi2019}), we expect that technology at this near-future stage should be capable of accurately recognizing users' voice and gesture with only a small margin of error. For example, the JHU-MIT consortium \cite{Villalba2020} achieved a low error rate of 1.53\% on the Speakers in the Wild (SITW) dataset, which includes voice samples from 300 users acquired across unconstrained conditions (e.g., stadiums, outdoors) \cite{McLaren2016TheSI}. Likewise, state-of-the-art gesture recognition systems \cite{Maghoumi2019} have achieved a maximum error rate of 8.7\% using multiple gesture recognition datasets (e.g., NTU RGB+D dataset \cite{Shahroudy2016} and UT Kinect dataset \cite{Xia2012}). Hence, a system at this near-future stage could now support a two-factor modality combination where natural modalities are both the first and second factors. Similar to N1, the use of two-factor makes this concept difficult to spoof, but its reliance on natural modalities still makes this concept observable. Therefore, users will still find this concept prone to risk of exposure (i.e., not private) and may feel awkward using this modality in the presence of others (i.e., not comfortable). Users could still find this concept unreliable as it may not perfectly recognize voice and gesture inputs in certain situations such as a multi-speaker scenario for voice recognition \cite{Villalba2020} or situations where certain body parts are occluded during gesture recognition \cite{Maghoumi2019}. This concept would also still not be easy to use since it requires extra effort from the user. However, in contrast to N1, users may find this concept easier to use since it only relies on modalities that are natural, which at this stage, users would have adequate experience using for authentication: \textit{+/- ease – comfort – reliability + safety – privacy}.
\paragraph{Far Future (N3):}
At this stage, the goal is to transition from N2 to true natural multimodal authentication while conforming to all facets of our design space. In the far future, we expect that technologies should have advanced to the extent that they are capable of recognizing voice and body gestures from the user with high accuracy, thus satisfying users’ mental models for reliability. Therefore, we could design a system that integrates natural multimodal authentication into the environment based on continuous authentication (i.e., periodically authenticating the user during interaction with the system \cite{Niinuma2010}). In smartphones, continuous authentication has been achieved using users’ behavioral patterns, such as gait, measured using accelerometers and gyroscopes, or touchscreen interactions \cite{Sitova2016}. A natural multimodal authentication system that uses continuous authentication would not require the user to actively interact with it to establish their identity before authentication can be performed, but rather passive authentication from observations of the users’ natural behaviors would be possible. For example, one possible design concept is to combine recognition of the user's voice or gesture with behavioral data (e.g., surface vibration, gait) measured through a smart floor. As the user walks and speaks in the environment, the system could continuously record their body vibrations and voice signals to authenticate the user. Another possible option is the design concept proposed by Feng et al. in which they measure a user’s body surface vibrations using wearables (e.g., eyeglasses, necklaces) and match it with the user’s voice signal \cite{10.1145/3117811.3117823} for continuous authentication. Since the authentication process is performed seamlessly based on users' natural interactions with the smart environment, the discreet nature of such an authentication mechanism solves the problem of observability that was associated with using natural modalities by our users. As a result, this potential design concept satisfies our users’ goal of being comfortable and easy to use because the user does not need to actively perform any gestures that may make them feel awkward and they can authenticate effortlessly. This concept also satisfies our users’ goal of being private (i.e., unobservable) and difficult to spoof since the authentication process is carried out behind the scenes on an ongoing basis. Thus, the proposed design concept has the potential to conform to all facets of our design space: \textit{+ ease + comfort + reliability + safety +privacy}.

These design concepts illustrate the potential of our design space lens to inspire design concepts that consider both the usability and security aspects of multimodal user authentication in smart environments. Some of these concepts may seem unrealistic now, but it is our hope that they can lead to future research and design into well-balanced user-centered authentication interactions for the future of smart environments.
\subsection{Limitations and Future Work}
Our study design has some limitations. Like other studies that recruit participants through Amazon Mechanical Turk, our participants were predominantly from the United States. Therefore, our findings may not be generalizable to other countries or cultures. Future work can redeploy our survey in other countries or cultures and investigate whether user authentication preferences differ across different cultures. Another limitation is that our survey duration, which was approximately one hour, could have increased participants’ fatigue, thus encouraging missing data. Although we did include an attention check question and used statistical tests to maintain the validity of our data, future research that may want to use our survey to understand user authentication preferences should consider shorter time durations (e.g., by splitting the questions). 
\section{CONCLUSION}
As users shift from interacting actively with screens to interacting seamlessly with smart technology that does not require screens, natural communication modalities such as voice and gesture become more suitable options for interaction and authentication. We presented data from 160 respondents (117 MTurkers and 43 CS students) to a comprehensive survey asking about authentication preferences for natural modalities for authentication (e.g., voice recognition), in contexts when others are present (e.g., family, co-workers) and for different usability metrics (e.g., ease, comfort). We found that users place less trust in natural authentication modalities compared to traditional authentication modalities (e.g., fingerprint scanner), so they would rather combine natural input modalities for two-factor authentication than solely using one of those modalities for authentication. We also found that, compared to traditional authentication modalities, users are not as willing to use natural authentication modalities except in the presence of people they trust, due to the risk of exposure and the awkwardness associated with using the modality. From our findings, we saw that users’ mental models center around how safe and comfortable these modalities are in public and private settings. We presented an analysis and exploration of the design space based on users’ current mental models for the future of secure and usable smart technology. Our work will inform the future design of natural multimodal authentication that users are willing to use for interaction with smart technology.

\section*{Acknowledgements}
We thank Discover Financial Services for their support of this work. Any opinions, findings, and conclusions or recommendations expressed in this paper are those of the authors and do not necessarily reflect these agencies’ views.

\bibliographystyle{unsrtnat}
\bibliography{references}  





\appendix
\section{Appendix: Online Survey Questionnaire}
\subsection{Demographics}
We first asked all participants common demographic questions including their age, gender, occupation, and language(s) spoken fluently. Then we asked about:
\begin{itemize}
    \item Income level: $\bigcirc \text{Less than \$20,000} \quad \bigcirc \text{\$20,000 to \$34,999} \quad  \bigcirc \text{\$35,000 to \$49,999} \quad  \bigcirc \text{\$50,000 to \$74,999} \quad  \bigcirc \text{\$75,000 to \$99,999} \quad  \bigcirc \text{\$100,000 to \$149,999} \quad  \bigcirc \text{\$150,000 to \$199,999} \quad  \bigcirc \text{\$200,000 or more}$
    \item Highest level of education completed: $\bigcirc \text{Less than high school} \quad  \bigcirc \text{High school degree or equivalent (e.g., GED)} \quad  \bigcirc \text{Some college but no degree} \quad  \bigcirc \text{Associate degree} \quad  \bigcirc \text{Bachelor degree} \quad  \bigcirc \text{Graduate degree} \quad  \bigcirc \text{Other(please specify)}$ \rule{3cm}{0.4pt}\hfill
    \item Experience with specific technologies Amazon Alexa/Microsoft Kinect/Touchscreen devices (e.g., smartphones, tablets)/Laptops/Other (please specify): $\bigcirc \text{Often} \quad  \bigcirc \text{Sometimes} \quad  \bigcirc \text{Rarely} \quad  \bigcirc \text{Never} \quad  \bigcirc \text{Never heard of it}$
\end{itemize}
\subsection{Methods used for authentication}
\begin{itemize}
    \item Have you ever had any of your accounts compromised where someone had access without your permission (e.g., email, Facebook, Twitter, bank info, computer account etc.)? $\bigcirc \text{Yes} \quad  \bigcirc \text{No}$
    \item Which authentication methods protected the account? $\bigcirc \text{Password/PIN} \quad  \bigcirc \text{Facial/Fingerprint} \quad  \bigcirc \text{Voice} \quad  \bigcirc \text{Body movement} \quad  \bigcirc \text{Pattern} \quad  \bigcirc \text{None} \quad  \bigcirc \text{Other(please specify)}$\rule{3cm}{0.4pt}\hfill
    \item Do you know which ones the attacker used to gain access? $\bigcirc \text{Yes} \quad  \bigcirc \text{No}$
    \item Which authentication methods were used to gain access? $\bigcirc \text{Password/PIN} \quad  \bigcirc \text{Facial/Fingerprint} \quad  \bigcirc \text{Voice} \quad  \bigcirc \text{Body movement} \quad  \bigcirc \text{Pattern} \quad  \bigcirc \text{None} \quad  \bigcirc \text{Other(please specify)}$\rule{3cm}{0.4pt}\hfill
\end{itemize}
\subsection{Accounts Compromised}
\begin{itemize}
    \item Have you ever had any of your accounts compromised where someone had access without your permission (e.g., email, Facebook, Twitter, bank info, computer account etc.)? $\bigcirc \text{Yes} \quad  \bigcirc \text{No}$
    \item Which authentication methods protected the account? $\bigcirc \text{Password/PIN} \quad  \bigcirc \text{Facial/Fingerprint} \quad  \bigcirc \text{Voice} \quad  \bigcirc \text{Body movement} \quad  \bigcirc \text{Pattern} \quad  \bigcirc \text{None} \quad  \bigcirc \text{Other(please specify)}$\rule{3cm}{0.4pt}\hfill
    \item Do you know which ones the attacker used to gain access? $\bigcirc \text{Yes} \quad  \bigcirc \text{No}$
    \item Which authentication methods were used to gain access? $\bigcirc \text{Password/PIN} \quad  \bigcirc \text{Facial/Fingerprint} \quad  \bigcirc \text{Voice} \quad  \bigcirc \text{Body movement} \quad  \bigcirc \text{Pattern} \quad  \bigcirc \text{None} \quad  \bigcirc \text{Other(please specify)}$\rule{3cm}{0.4pt}\hfill
\end{itemize}
\subsection{Password forgotten}
\begin{itemize}
    \item Have you ever forgotten a password or other authentication method? $\bigcirc \text{Yes} \quad  \bigcirc \text{No}$
    \item What authentication method and how did you recover it?\rule{3cm}{0.4pt}\hfill
\end{itemize}
\subsection{Multiple-Answer}
This is a question for all participants.
\begin{itemize}
    \item Please check which of these categories you would prefer to require authentication before providing access. Think about the possibility of someone else being able to perform the tasks in the category. Assume each category is independent of the state-of-the-art device/platform within which the scenario exists.
    \begin{itemize}
        \item Calendar: $\square \text{Checking an event time.} \quad \square \text{Creating/Editing an event.} \quad \square \text{Sharing an event.} \\ \quad \square \text{Deleting an event.}$
        \item Email/Messages:  $\square \text{Checking/Reading emails.} \quad \square \text{Creating emails.} \quad \square \text{Deleting emails.}$
        \item Online Banking: $\square \text{View Balance.} \quad \square \text{View Transaction.} \quad \square \text{Deposit.} \quad \square \text{Transfer.} \quad \square \text{Pay Bill.}$
        \item Photos: $\quad \square \text{View photos.} \quad \square \text{Share photos.} \quad \square \text{Delete photos.}$
        \item Contacts: $\square \text{View contacts.} \quad \square \text{Share contacts.} \quad \square \text{Delete contacts.} \quad \square \text{Create contacts.}$
        \item Making a purchase:
        $\square \text{Adding items to the cart.} \quad \square \text{Entering shipping information.} \\ \quad \square \text{Entering payment information.} \quad \square \text{Cancelling an order.}$
        \item Access online information: $\square \text{Look up online articles.} \quad \square \text{Look up browser account information.} \\ \quad \square \text{Download information.}$
        \item Are there any other scenarios that you would prefer to require authentication that are not listed? (please specify)\rule{3cm}{0.4pt}\hfill
    \end{itemize}
\end{itemize}
\subsection{Scenario Authentication Questions - Calendar / Email or Messages / Online Banking / Photos / Contacts / Making Purchases / Access Online}
\begin{figure}[h]
	\centering
        \includegraphics[scale=0.7]{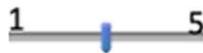}
	\caption{The slider for the participant to rate from 1 to 5.}
	\label{fig:fig4}
\end{figure}

Question for only participants who preferred to authenticate for a given category in the previous question. Also, for each item we provided the slider shown in Figure \ref{fig:fig4} for the participant to rate.
\begin{itemize}
    \item You indicated that you would prefer to require authentication before using your calendar / banking online / accessing your photos / accessing your contacts / making purchases online / accessing online information. Rate the likelihood that you would be willing to accept the following type of authentication to verify your identity. Rate from 1 to 5 where 1 is very unlikely and 5 is very likely:
    \begin{itemize}
        \item Type a password
        \item Use voice recognition
        \item Make a body movement in front of a camera
        \item Use a fingerprint scanner
        \item Use face recognition
        \item Other (please specify)
    \end{itemize}
\end{itemize}
\subsection{General Authentication Questions}
This is a question for all participants. Also, for each item we provided the slider shown in Figure \ref{fig:fig4} for the participant to rate.
\begin{itemize}
    \item Imagine a situation where you are alone or with only the listed people present. Rate the likelihood that you would be willing to accept typing a password / voice recognition / making body movements / using a fingerprint scanner / using face recognition to verify your identity from 1 to 5 where 1 is very unlikely and 5 is very likely:
    \begin{itemize}
        \item Alone
        \item Significant other
        \item Roommate
        \item Close friends
        \item Acquaintances
        \item Family
        \item Co-workers
        \item General Public
        \item Briefly explain the reason for the context with the highest rating:\rule{3cm}{0.4pt}\hfill
        \item Briefly explain the reason for the context with the lowest rating:\rule{3cm}{0.4pt}\hfill
    \end{itemize}
\end{itemize}
\subsection{Ease of Use / Naturalness / Security / Comfortable}
This is a question for all participants. We asked participants to briefly explain their answers for each question. Also, for each item we provided the slider shown in Figure \ref{fig:fig4} for the participant to rate.
\begin{itemize}
    \item Rate how easy / natural or intuitive / secure / comfortable you feel using the following type of authentication to verify your identity is from 1 to 5 where 1 is not very easy to use and 5 is very easy to use:
    \begin{itemize}
        \item Type a password
        \item Use voice recognition 
        \item Make a body movement in front of a camera (e.g., wave, thumbs up)
        \item Use a fingerprint scanner 
        \item Use face recognition
    \end{itemize}
\end{itemize}
\subsection{Two Factor Authentication}
This is a question for all participants. We covered all the combinations in 5 questions. Each question has one less option than the previous one, where one of the combinations was already covered. Also, for each item we provided the slider shown in Figure \ref{fig:fig4} for the participant to rate.
\begin{itemize}
    \item Imagine you have a system that allows for two-factor authentication (i.e., using a combination of two authentication methods). Rate how likely you are to use a typed password / biometric / text message / phone call / body movement with the following authentication methods from 1 to 5 with 1 being very unlikely and 5 being very likely:
    \begin{itemize}
        \item Biometric (fingerprint scanner, face recognition)
        \item Text message
        \item Phone call
        \item Body movement
        \item Voice recognition
        \item Briefly explain your response for the highest rated authentication method:\rule{3cm}{0.4pt}\hfill
        \item Briefly explain your response for the lowest rated authentication method:\rule{3cm}{0.4pt}\hfill
    \end{itemize}
\end{itemize}
\subsection{Attention Check}
This is a question for all participants. This question was used to determine is participants had read and answered the questions, instead of clicking responses randomly.
\begin{itemize}
    \item You have almost completed the survey. We have to make sure that our data are valid and not biased. Specifically, we are interested in whether you read the instructions closely. Please select the option 'no answer' for this question. How long did you feel this survey was? $\bigcirc \text{Very long} \quad \bigcirc \text{Long} \quad \bigcirc \text{Neither short or long} \quad \bigcirc \text{Very short} \quad \bigcirc \text{No answer}$
\end{itemize}
\section{Appendix: Qualitative Code Definitions}
\begin{longtable}{l p{0.8\linewidth}}
    \toprule
    \textbf{Code} & \textbf{Definition} \\
    \midrule
    \endhead
    \textbf{Attention} & User is conscious about security and does things with authentication to ensure security. (e.g., I use a complicated password for my accounts.) \\
    \\
    \textbf{Comfort} & This code generally relates to the comfort of the person or others around them, which also includes disruptive/ bothersome to others. In other words, the authentication mode makes the person or others comfortable/uncomfortable or feels awkward/not awkward. (e.g., I feel awkward using body movements.) \\
    \\
    \textbf{Comparison} & This code relates to comparison between modalities. (e.g., Typing is easier than making a body movement. Body movement seems a lot less consistent than face recognition.) \\
    \\
    \textbf{Constraint} & The code relates to factors that hinders the use or the effectiveness of the authentication method, such as environmental, physical, device, methodological, monetary cost. Additionally, personal characteristics that encourages or discourages use of the modality is also included. (e.g., I wear gloves in the winter, so it is hard to do fingerprint recognition.) \\
    \\
    \textbf{Control} & This code covers the cases where the user states that they do not want anyone seeing information (without permission). (e.g., I only want myself to have access. I do not want anyone to access it without my authorization.) \\
    \\
    \textbf{Ease of Use} & This code relates to how easy or difficult the method is, and the effort required to remember, perform, or set up the authentication. (e.g., It is just pushing on a screen; I do that all the time.) \\
    \\
    \textbf{Frequency} & The code refers to the frequency of the action that is performed. It also includes the cases where the action is performed multiple times, the action is slowed down, or the action is not being performed at all. (e.g., I do this all the time. I do this a lot.) \\
    \\
    \textbf{Last Resort} & This code refers to the case where the user says that they would use this authentication mode only if they had no other options. (e.g., If I had no other choice, I guess I would have to do voice because I need to authenticate.) \\
    \\
    \textbf{Permanence} & This code relates to whether the action is reversible or not, and whether the information can be recovered or not after the action has taken place. (e.g., I can delete randomly created contacts.) \\
    \\
    \textbf{Preference} & This code refers to any form of bias towards or against the method, which means the willingness/unwillingness to perform the authentication due to factors such as familiarity or prior experience with the authentication mode. It also includes novelty. (e.g., I do not like face recognition. I would not want to use body movement.) \\
    \\
    \textbf{Privacy} & This code relates to sensitivity of the information. (e.g., I feel like it is my duty to protect the personal information of people in my contacts.) \\
    \\
    \textbf{Redundant} & This code relates to the case where the two authentication methods together make no difference either because they are very similar or because they are pointless together. (e.g., Phone calls are already voice-based, this would be redundant.) \\
    \\
    \textbf{Reliability} & This code refers to the reliability of the authentication method with regards to accuracy/consistency of being able to always authenticate the person. (e.g., Face recognition can be faulty sometimes.) \\
    \\
    \textbf{Risk} & This code relates to the case where the user thinks there is no repercussion or when the user feels there is no risk. It represents the presence of repercussions that are involved with access to a task, and must be related to an action. (e.g., Embarrassing photos leaked, why would you want to hide that? I do not care if someone creates one.) \\
    \\
    \textbf{Security} & This code relates to the security and safety of the authentication method with regards to how difficult it is to manipulate, how easy it is to spoof, or how discreet it is. (e.g., Voice recognition is semi-secure. Types password would be the least intrusive.) \\
    \\
    \textbf{Time} & The code relates to the speed, efficiency, or time it takes to perform an action. (e.g., Fingerprint scanner is fast.) \\
    \\
    \textbf{Trust} & The code relates to the case where the context is rated because of trusting/distrusting people or being familiar/unfamiliar with people. (e.g., I do not have an issue unlocking with face recognition around these people as I trust them.) \\
    \\
    \textbf{Uniqueness} & This code relates to the case where the individual uses something unique to authenticate. (e.g., I am the only one with this fingerprint. Since no one can copy my face I would be comfortable with anybody.) \\
    \\
    \textbf{Other} & This code covers the cases that do not fall under any other code. \\
    \bottomrule
\end{longtable}

\end{document}